\documentclass[a4,aps,amsmath,floatfix,nofootinbib,10pt]{revtex4}
\usepackage{graphicx}
\usepackage{color}
\usepackage{enumerate}
\newcommand{\be}{\begin{equation}}
\newcommand{\ee}{\end{equation}}
\newcommand{\ba}{\begin{array}}
\newcommand{\ea}{\end{array}}
\newcommand{\bea}{\begin{eqnarray}}
\newcommand{\eea}{\end{eqnarray}}
\newcommand{\bdm}{\begin{displaymath}}
\newcommand{\edm}{\end{displaymath}}

\begin{document}


\title{Hysteresis in Random-field Ising model on a \\ Bethe lattice with 
a mixed coordination number}

\author{Prabodh Shukla and Diana Thongjaomayum}

\affiliation{Physics Department \\ North Eastern Hill University \\ 
Shillong-793 022, India}

\begin{abstract}

We study zero-temperature hysteresis in the random-field Ising model on 
a Bethe lattice where a fraction $c$ of the sites have coordination 
number $z=4$ while the remaining fraction $1-c$ have $z=3$. Numerical 
simulations as well as probabilistic methods are used to show the 
existence of critical hysteresis for all values of $c > 0$. This extends 
earlier results for $c=0$ and $c=1$ to the entire range $0 \le c \le 1$, 
and provides new insight in non-equilibrium critical phenomena.

\end{abstract}

\maketitle

\section{Introduction}

The path-breaking study of zero-temperature hysteresis in the 
random-field Ising model ~\cite{sethna1, sethna2, dahmen, perkovic, 
sethna3}, has enhanced our understanding of a complex system's response 
to a slowly varying applied field. It explains several features observed 
in experiments; hysteresis, Barkhausen noise, return point memory, 
discontinuity in magnetization , and a non-equilibrium critical point. 
The non-equilibrium critical point is accompanied by anomalous 
scale-invariant fluctuations (avalanches) akin to those observed in the 
vicinity of an equilibrium second order phase transition. Consequently, 
the non-equilibrium critical point shows many of the same universal 
features as the equilibrium one. However, there appears to be a 
difference when it comes to the role of a lower critical dimension 
$d_l$. If the dimension of the system $d$ is lower than $d_l$, 
equilibrium thermal fluctuations are too large to allow a phase 
transition to an ordered state. For $d>d_l$, the system can make a phase 
transition if its temperature drops below a critical temperature. In the 
equilibrium case, $d_l=1$ for the Ising model, and $d_l=2$ for the 
random-field Ising model~\cite{imry-ma,aizenman}. For the 2d Ising model 
solved by Onsager ~\cite{onsager} on a square lattice, the existence of 
a critical point is not supposed to depend on whether the lattice is 
square, triangular, or honeycomb. The short range structure of the 
lattice is irrelevant under a diverging correlation length. It is not 
unreasonable to expect the same for the non-equilibrium critical point. 
However, this is not borne out by numerical studies of the random-field 
Ising model, and our understanding of the general conditions for the 
existence of a non-equilibrium critical point remains far from 
satisfactory.

In the non-equilibrium random-field Ising model at zero temperature, with 
the random-field having mean value zero and standard deviation $\sigma$, 
$\sigma$ plays a role analogous to temperature in the equilibrium model.  
Numerical work on a simple cubic lattice shows that there is a critical 
value $\sigma_c$ such that the response of the system to a steadily 
increasing field $h$ has a discontinuity at $h=h_c$ if $\sigma < 
\sigma_c$. The size of the discontinuity decreases with increasing 
$\sigma$ and reduces to zero at a critical point; $\sigma=\sigma_c$, 
$h=h_c$. For $\sigma > \sigma_c$ the response is smooth and free from 
any singularity. There is no singularity in $d=1$ for any value of 
$\sigma$ ~\cite{shukla}. For $d=2$, it was initially unclear if there is 
or is not a critical point on a square lattice ~\cite{perkovic,sethna3}. 
It suggested that $d_l$ may be equal to 2 as in the case of the thermal 
equilibrium random-field Ising model. However, this is not true. There 
is evidence that the existence of critical hysteresis depends on the 
coordination number of the lattice rather than the dimensionality of 
space in which the lattice is embedded. A large-scale numerical 
simulation shows that critical hysteresis is present on a square 
lattice~\cite{spasojevic}. It is also present on a triangular lattice 
~\cite{diana} but absent on a honeycomb lattice~\cite{sabhapandit}. An 
exact solution of the model on a Bethe lattice of coordination number 
$z$ reveals that the critical point exists only on lattices with $z \ge 
4$ ~\cite{dhar}. The significance of this result seems to extend beyond 
a Bethe lattice. Numerical work shows the absence of a critical point on 
periodic lattices with $z=3$ and its presence on lattices with $z \ge 4$ 
irrespective of the dimensionality of the lattice ~\cite{sabhapandit}.

A question arises as to whether a lattice with a fractional coordination 
number $4-\epsilon$ ($0 < \epsilon <1$) can support critical hysteresis. 
This question was examined in reference ~\cite{kurbah}. Starting from a 
triangular lattice ($z=6$), a fraction $f$ of the sites on one of its 
three sub-lattices were removed gradually and randomly till the lattice 
reduced to a honeycomb lattice ($z=3$). Earlier work had established the 
presence of critical hysteresis on the triangular lattice ($f=0$), and 
its absence on the honeycomb lattice ($f=1$). Numerical work for $f > 0$ 
indicated that the critical hysteresis disappears if $f>2/3$, i.e. if 
the effective coordination number $z_{eff}=3(2-f)$ of each of the two 
undiluted sub-lattices of the triangular lattice drops below $z_{eff}=4$ 
~\cite{kurbah}. At $z_{eff}< 4$, the probability of a spanning path 
through occupied sites on the triangular lattice goes to zero. However, 
one can construct other lattices with $3<z_{eff}<4$ which have spanning 
clusters across the lattice. This motivates us to reexamine the question 
on a Bethe lattice of a mixed coordination number: a fraction $c$ of the 
sites have $z=4$ nearest neighbors, and the remaining fraction $1-c$ 
have $z=3$ nearest neighbors. We study the lattice with the mixed 
coordination number numerically as well as analytically. The numerical 
work is performed on a random graph, but drawing conclusions from it 
regarding the existence of a critical point is as tedious as in the case 
of the periodic lattice. Fortunately, with the benefit of an analytic 
solution of the problem, it becomes easier to understand the numerical 
work. Our conclusion is that the critical hysteresis is present for all 
values of $c >0$.

\section{The model, simulations, and data analysis}

The Hamiltonian for the random-field Ising model with interaction $J >0$ 
between nearest neighbor sites $i$ and $j$ is,

\be H=-J\sum_{i,j}{s_{i}s_{j}}-\sum_{i}h_{i}s_{i}-h\sum_{i}s_{i}\ee

Here $s_i=\pm{1}$ is an Ising spin, $h_i$ is a random-field, and 
$h$ a uniform external field. The random-field has a Gaussian 
distribution with mean value zero and standard deviation $\sigma$. 

The spin at time $t+1$ is updated by aligning it along the local field 
$l_i$ at site $i$ at time $t$;

\be s_i(t+1)=sign\mbox{ } l_i(t);\mbox{ } l_i=J \sum_{j}s_{j}+h_{i}+h 
\ee

Simulations are performed on a random graph of $N$ sites where $c N$ 
sites have $z=4$ nearest neighbors, and the rest have $z=3$. A random 
graph for $z=4$ is constructed as described in reference ~\cite{dhar} 
and then a fraction of fourth neighbor bonds are removed. Figure (1) 
illustrates a random graph of 12 sites with 4 sites having $z=4$ and 8 
sites having $z=3$. The actual simulations are performed on graphs of 
size $N > 10^6$, and $0 \le c \le 1$. We generate a quenched 
random-field distribution $\{h_i\}$ for a fixed value of $c$ and 
$\sigma$, and start with a sufficiently negative value of $h$ when all 
spins are down $\{s_i=-1\}$. The applied field $h$ is then increased 
slowly till some site becomes unstable i.e. it sees a positive local 
field at its site. At this point, $h$ is kept fixed and the system is 
updated iteratively till a fixed point is reached i.e.  each spin is 
aligned along the local field at its site. The spins that flip up on the 
way to the fixed point form a connected cluster, and the number of spins 
that flip up is the size of the avalanche at $h$. Now $h$ is increased 
to the next instability in the system, and again the size of the 
avalanche is calculated as above. The process is repeated till all spins 
are up and stable. The locus of the fixed points gives the magnetization 
curve $m(h;\sigma,c)$ in increasing applied field $h$. The magnetization 
curve is macroscopically smooth but noisy (Barkhausen noise) at a 
microscopic scale because of the avalanches that separate neighboring 
fixed points. Our objective is to determine if the magnetization curve 
has a discontinuity i.e. if two neighboring fixed points are separated 
by a macroscopic avalanche.

The value of $h$ where a discontinuity occurs in $m(h;\sigma,c)$, if 
indeed there is a discontinuity, shifts slightly from one configuration 
of random fields to another for the same size of the system. Averaging 
the magnetization over different configurations tends to smoothen the 
curve and hide the discontinuity. A discontinuity is better seen in a 
single run of a large system. Figure (2) shows magnetization curves for 
a single run for $N=10^8$, $\sigma=0.5$ and different values of $c$. We 
choose $\sigma=0.5$ because it is known from earlier work that 
$m(h;\sigma=0.5,c)$ is continuous for $c=0$ but discontinuous for $c=1$. 
Therefore for $\sigma=0.5$, there must be a value of $c$ where the 
behavior changes from smooth to discontinuous. However, it is difficult 
to read this from figure (2). Each curve in figure (2) seems to have a 
discontinuity, although the curve for $c=0$ has a slightly different 
character. With the benefit of an exact solution for $c=0$ we know that 
the curve will become smooth as the system size is increased beyond 
$N=10^8$. However, deciding a discontinuity by visual inspection is 
inadequate, and particularly so for locating the critical value 
$\sigma_c$ above which the discontinuity may disappear.

Another method of analyzing the data is to use the distribution of 
avalanche sizes ~\cite{farrow}. The probability $P(s;\sigma,c)$ of an 
avalanche of size $s$ on the trajectory $m(h;\sigma,c)$ is generally a 
product of two terms: (i) one that decreases exponentially with 
increasing $s$ and represents microscopic avalanches, and (ii) a delta 
function peak at a very large value of $s$ representing a macroscopic 
discontinuity. Figure (3) shows a plot of $\ln{P(s;\sigma=0.5,c)}$ $vs.$ 
$s$ for $c=0, 0.5, 1$ respectively. Figure (3) presents the data for 
$N=10^6$, averaged over $10^4$ configurations of the random-field 
distribution. An exact solution for integer $c$ predicts that 
$m(h;\sigma=0.5,c)$ is continuous for $c=0$, but discontinuous for 
$c=1$. The motivation behind the plot in figure (3) is to get an 
indication if the case of $c=0.5$ is more like $c=0$ or $c=1$. Evidently 
no definite conclusion can be reached in this regard from figure (3). 
The curve for $c=0.5$ seems to have a mixture of features of $c=0$ and 
$c=1$; there is an indication of a delta function peak, but there is 
also a relatively large proportion of small size avalanches which is 
characteristic of a smooth curve. It is fair to say that figure (3) 
alone does not give a clear indication of the nature of singularity for 
$c=0$ or $c=1$, let alone for intermediate values of $c$.

Evidently, finite size effects make it difficult to distinguish between 
a sharply rising continuous curve and one with a discontinuity. There 
does not appear to be a numerical method that can differentiate between 
these two types of curves with certainty. We also tested the ratio $R$ 
of the largest to the second largest avalanche on the magnetization 
curve. $R$ should diverge if a discontinuity is present, and therefore 
it may separate curves with a discontinuity from those without it. The 
efficacy of this method is rather poor for $\sigma=0.5$ but improves 
around $\sigma=1$ where it indicates the presence of a discontinuity for 
all positive values of $c$. However, we omit this analysis here and 
present an analytic solution of the problem which makes the situation 
clear.

\section{Analytic results}

As discussed in reference ~\cite{dhar}, the thermodynamic limit of a 
random graph has the same structure as the deep interior of a Cayley 
tree. Theoretical analysis is simpler on a Cayley tree because of the 
absence of loops on it. We are able to adapt the method used in 
reference ~\cite{dhar} to the present case. A Cayley tree with $z=4$ is 
shown schematically in figure (3). Initially all spins are kept down. 
Spins are relaxed starting from the surface of the tree and moving one 
level at a time towards the root of the tree. Thus spins at level $n+1$ 
are relaxed keeping their nearest neighbor at level $n$ down. Consider a 
particular site at level $n$. It has three neighbors at level $n+1$ and 
one at level $n-1$. Any one of its four neighbors may be missing with 
probability $1-c$. If the missing neighbor is at level $n-1$, the site 
in question forms the vertex of a sub-tree which is disconnected from 
the rest of the tree and therefore does not affect the root of the tree. 
However, in writing a recursion relation for the relaxation process, we 
find it convenient to assume that the missing neighbor lies at level 
$n+1$ only. It amounts to overestimating the presence of $z=4$ sites on 
the lattice but we correct for it later.

Now we focus on the three sites at level $n+1$ referred above. One of 
these may be missing with probability $1-c$. If not missing, it may be a 
$z=4$ site or a $z=3$ site. Let $P^{n+1}_4(h)$, and $P^{n+1}_3(h)$ be 
the probability that the spin is up in the two cases respectively. The 
average probability that the spin at the site is up is equal to $ 
<P^{n+1}(h)>=cP^{n+1}_4(h)+(1-c)P^{n+1}_3(h)$. It is easy to 
see that $P^{n}_4(h)$ and $P^{n}_3(h)$ are given by the recursion 
relations,

\bea P^{n}_4(h)=<P^{n+1}(h)>^{3}p_{43}(h)+3 
<P^{n+1}(h)>^{2}[1-<P^{n+1}(h)>]p_{42}(h) \nonumber \\
          +3 <P^{n+1}(h)>[1-<P^{n+1}(h)>]^{2}p_{41}(h) 
+ [1-<P^{n+1}(h)>]^{3}p_{40}(h)\eea
	        
\bea P^{n}_3(h)=<P^{n+1}(h)>^{2}p_{32}(h)+2<P^{n+1}(h)> 
[1-<P^{n+1}(h)>]p_{31}(h)\nonumber \\ +[1-<P^{n+1}(h)>]^{2}p_{30}(h)\eea

Here, $p_{zm}(h)$, $z=3,4$ and $m\le z$, is the probability that the 
random-field at a site with $z$ neighbors is large enough so that the 
spin at the site can flip up if $m$ of its neighbors are up.

\bdm p_{zm}(h)=\frac{1}{\sqrt{2 \pi \sigma^2}}\int_{(z-2 m)J-h}^\infty 
dh_i \exp{-(h_i^2/2 \sigma^2)} \hspace{.5cm} (m \le z) \edm

Equations (3) and (4) lead to fixed points $P^*_4(h)$ and $P^*_3(h)$ in 
the limit $n \rightarrow \infty$. Given that the spin at the root of the 
tree (i.e. any site in the deep interior of the Cayley tree) is down, 
the probability that its neighbor is up is given by, $P^*(h) =c P^*_4(h) 
+ (1-c) P^*_3(h)$. The probability that the spin at the root is up 
depends on whether the root is a $z=4$ site or a $z=3$ site, and is 
given respectively by the following equations.

\bea p_4(h)=[P^*(h)]^4 p_{44}(h)+4 [P^*(h)]^3 [1-P^*(h)] p_{43}(h)
             +6 [P^*(h)]^2 [1-P^*(h)]^2 p_{42}(h) \nonumber \\
             +4 [P^*(h)] [1-P^*(h)]^3 p_{41}(h)+ [1-P^*(h)]^4 p_{40}(h) \eea

\bea p_3(h)=[P^*(h)]^3 p_{33}(h)+3 [P^*(h)]^2 [1-P^*(h)] p_{32}(h)
             +3 P^*(h) [1-P^*(h)]^2 p_{31}(h) \nonumber \\
             +[1-P^*(h)]^3 p_{30}(h)\eea

The probability that the spin at the root is up is equal to $p(h)=c 
p_4(h) + (1-c) p_3(h)$, and the magnetization per site is equal to 
$m(h;\sigma,c)=2 p(h) -1$. The magnetization per site on $z=3$ and $z=4$ 
sites is given by $m_3(h;\sigma,c)=2 p_3(h) -1$ and $m_4(h;\sigma,c)=2 
p_4(h) -1$ respectively.

The discontinuity in magnetization is related to a discontinuity in the 
fixed points $P^*_4(h)$ and $P^*_3(h)$ as a function of $h$. Since 
$P^*_4(h)$ and $P^*_3(h)$ are determined by coupled equations, a 
discontinuity in $P^*_4(h)$ at $h$ is accompanied by a discontinuity in 
$P^*_3(h)$ as well. We focus on their average value $P^*(h) =c P^*_4(h) 
+ (1-c) P^*_3(h)$. Equations (3) and (4) reveal a critical value 
$\sigma_c$ for each concentration $c$ of $z=4$ sites: $\sigma_c$ 
separates discontinuous from continuous behavior of $P^*(h)$ and 
$m(h;\sigma,c)$. If $\sigma < \sigma_c$, $P^*(h)$ has three roots in a 
window of applied field centered at $h=J$. One of these roots is 
$P^*(h)=0.5$ at $h=J$, but this root is unstable. The other two roots 
are stable and correspond to a discontinuity in $P^*(h)$ which jumps up 
from a value $P^*(h) < 0.5$ to a higher value $P^*(h) > 0.5$. This 
corresponds to a jump in magnetization from a negative to a positive 
value. The size of the jump reduces as $\sigma$ increases, and vanishes 
at a non-equilibrium critical point at $\sigma=\sigma_c$ and $h=J$. At 
the critical point the two stable roots for $P^*(h)$ merge into each 
other. We determine $\sigma_c$ algebraically by requiring the two stable 
roots of the fixed point equation become a double root. The result is 
depicted in figure (5). It predicts that the discontinuity in $m(h;\sigma,c)$ 
occurs for all values of $c$ greater than zero and $\sigma < \sigma_c$; 
$\sigma_c$ decreases with decreasing $c$. For $\sigma > \sigma_c$, the 
magnetization curve is smooth.

Before comparing the theoretical result with numerical simulations, we 
make a correction to which we alluded earlier in this section. The 
recursion relations on the Cayley tree assume that a $z=3$ site at level 
$n$ is necessarily connected to its neighbor at level $n-1$. This 
overestimates the concentration of $z=4$ sites at level $n-1$ by a 
fraction $(1-c)/4$, i.e. the fraction of $z=4$ sites at level $n-1$ 
becomes $c(1+(1-c)/4)$. The correction propagates all the way to the 
deep interior of the tree. The effective concentration of $z=4$ sites in 
the vicinity of the central site becomes $c_{eff}=c(1+r+r^2+r^3+\ldots)$ 
where $r=(1-c)/4$. Summing the geometric series we get, $c_{eff} = 
4c/(3+c)$. Thus simulations on a random graph with a fraction $c$ of 
$z=4$ sites should match the theoretical result for $c_{eff}$. This is 
indeed the case. A few selected comparisons are shown in figure (6) and 
figure (7). Figure (6) shows magnetization $m(h;\sigma,c)$ for 
$c=0.1,\sigma=0.9$ and $c=0.9,\sigma=1.6$ near a discontinuity. 
Corresponding theoretical expressions have been superposed on the 
simulation results and the two fit each other quite well. The curve with 
the discontinuity closer to $h=J$ is for $c=0.1$, and the other for 
$c=0.9$. The critical values of $\sigma$ for $c=0.1$ and $c=0.9$ are 
$\sigma_c=0.93$ and $\sigma_c=1.69$ respectively. Since the simulations 
are for $\sigma < \sigma_c$ but close to $\sigma_c$, they show 
discontinuity at $h > J$ but close to it. Figure (7) shows 
$m_3(h;\sigma,c)$ and $m_4(h;\sigma,c)$, the magnetization per site on 
$z=3$ and $z=4$ sites respectively for $c=0.75$ and $\sigma=1.5$ 
($\sigma_c=1.56$). Again the corresponding theoretical expressions have 
been superimposed on the simulations and the fit is quite good as may be 
expected.

\section{Discussion}

The work presented above extends the treatment of critical hysteresis on 
Bethe lattices of integer coordination number to lattices with a 
fractional coordination number. It is significant from two points of 
view. The first point concerns the question of a lower critical 
coordination number $z_l$ vs. a lower critical dimension $d_l$. The 
existence of a non-equilibrium critical point is decided by $z_l$ rather 
than $d_l$. Earlier work suggested $z_l=4$, but the present result shows 
$z_l > 3$. The physical significance of $z_l$ is not very clear at 
present. Mathematically, a discontinuity in $m(h;\sigma,c)$ occurs when 
$m(h;\sigma,c)$ has an "$s$-shape". An $s$-shape requires three 
solutions for $m(h;\sigma,c)$ for the same value of the applied field 
$h$ in some range of $h$. The middle part of the $s$-shaped curve on 
which $m(h;\sigma,c)$ decreases with increasing $h$ is physically 
unstable causing the magnetization to jump over it. Earlier work 
examined only integer values of $z$ and found $z=4$ to be the smallest 
value of $z$ for which $m(h;\sigma,c)$ could have three solutions. 
Somewhat surprisingly, this result for the Bethe lattice seems to apply 
also to several periodic lattices of integer coordination number, 
irrespective of the dimensionality of space $d$ in which the lattice is 
embedded. We have examined fractional coordination numbers and find $z_l
> 3$. Of course, the coordination number $z$ of a lattice
site is necessarily an integer. We constructed random graphs with a 
mixture of sites with $z=3$ and $z=4$ so that the average coordination 
number $z$ lies between $3$ and $4$. Random graphs are excellent 
representations of a Bethe lattice or the deep interior of a Cayley 
tree. We studied the problem numerically on random graphs, but 
analytically on a Cayley tree. The numerical results fit corresponding 
theoretical expressions quite well. After an analytic solution is found, 
one may think the role of numerical work is essentially to check the 
solution. However, we have presented a brief account of our numerical 
effort to emphasize its difficulty in deciding the question of a true 
discontinuity in the magnetization. Methods which work on lattices of a 
uniform coordination number become less efficient when $z$ is 
disordered. Even on random graphs, it is hard to draw clear conclusions 
unless aided by analytical results. Since it is extremely difficult to 
obtain exact solutions on periodic lattices, search for better numerical 
methods has to continue.

The second point is that lattices with a mixed coordination number are 
quite common and important in the field of amorphous solids 
~\cite{mitra}. In magnetism, the coordination number of a site 
determines the exchange and the anisotropy field at the site and 
therefore the predominant nature of the spin at the site i.e. discrete 
or continuous. This in turn affects the relaxation rate of the spin, and 
the shape of the hysteresis loop ~\cite{kharwanlang}. Similar effects 
are also important in molecular magnetism and its industrial 
applications ~\cite{zadrozny}. Thus the work presented here may be 
useful in understanding non-equilibrium phase transitions in a wider 
class of disordered materials and their applications.

\begin{figure}[p] 
\includegraphics[width=0.75\textwidth,angle=0]{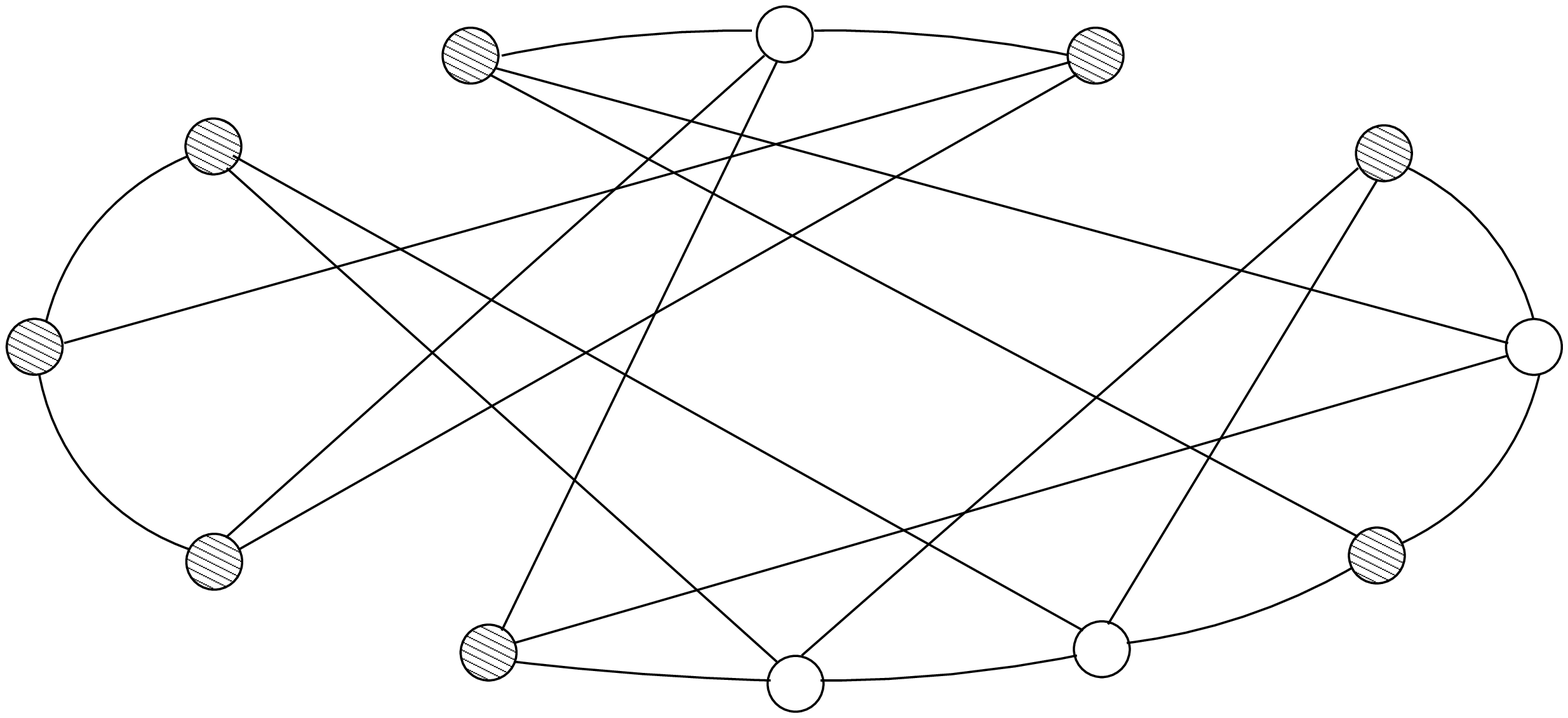} \caption{A 
random graph with 4 sites (open circles) having $z=4$, and 8 sites (filled 
circles) having $z=3$.} \label{fig1} \end{figure}

\begin{figure}[p] 
\includegraphics[width=0.5\textwidth,angle=0]{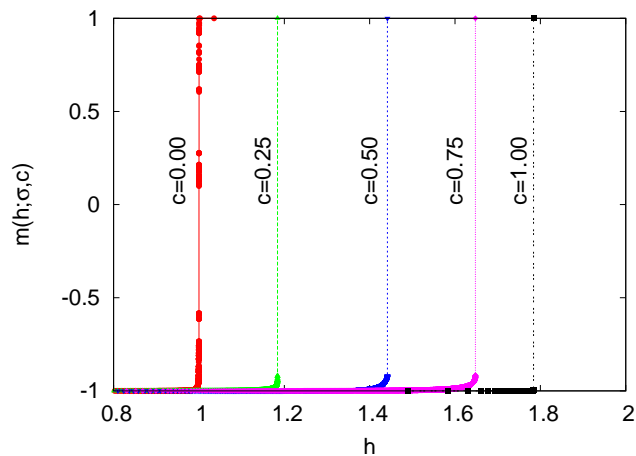} 
\caption{Magnetization $m(h;\sigma,c)$ in increasing field $h$ for 
$N=10^8$, $\sigma=0.50$ and values of $c$ as marked vertically next to 
each curve. } \label{fig2} \end{figure}

\begin{figure}[p] 
\includegraphics[width=0.5\textwidth,angle=0]{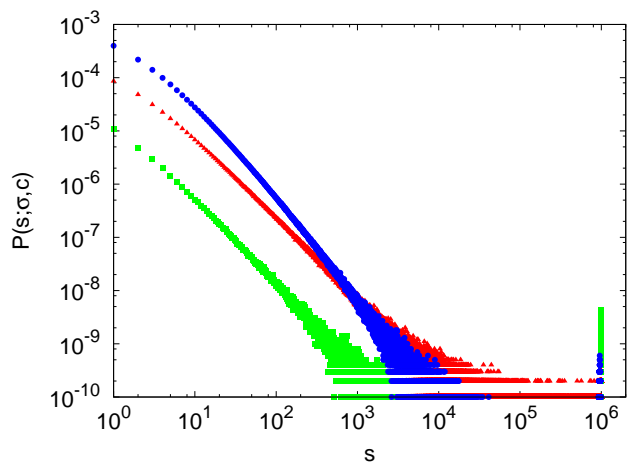} 
\caption{Probability $P(s;\sigma,c)$ of an avalanche of size $s$ along 
the lower half of the hysteresis loop for $\sigma=0.50$. The data is 
obtained from avalanches in a system of size $N=10^6$, averaged over 
$10^4$ independent configurations of the random field. The figure shows 
three cases (i) red (triangles) for $c=0$, (ii) blue (circles) for 
$c=0.5$, and (iii) green (squares) for $c=1$. Theory predicts 
discontinuous magnetization curves for $c=0.5$ and $c=1$, but continuous 
for $c=0$.} \label{fig3} \end{figure}

\begin{figure}[p] 
\includegraphics[width=0.5\textwidth,angle=0]{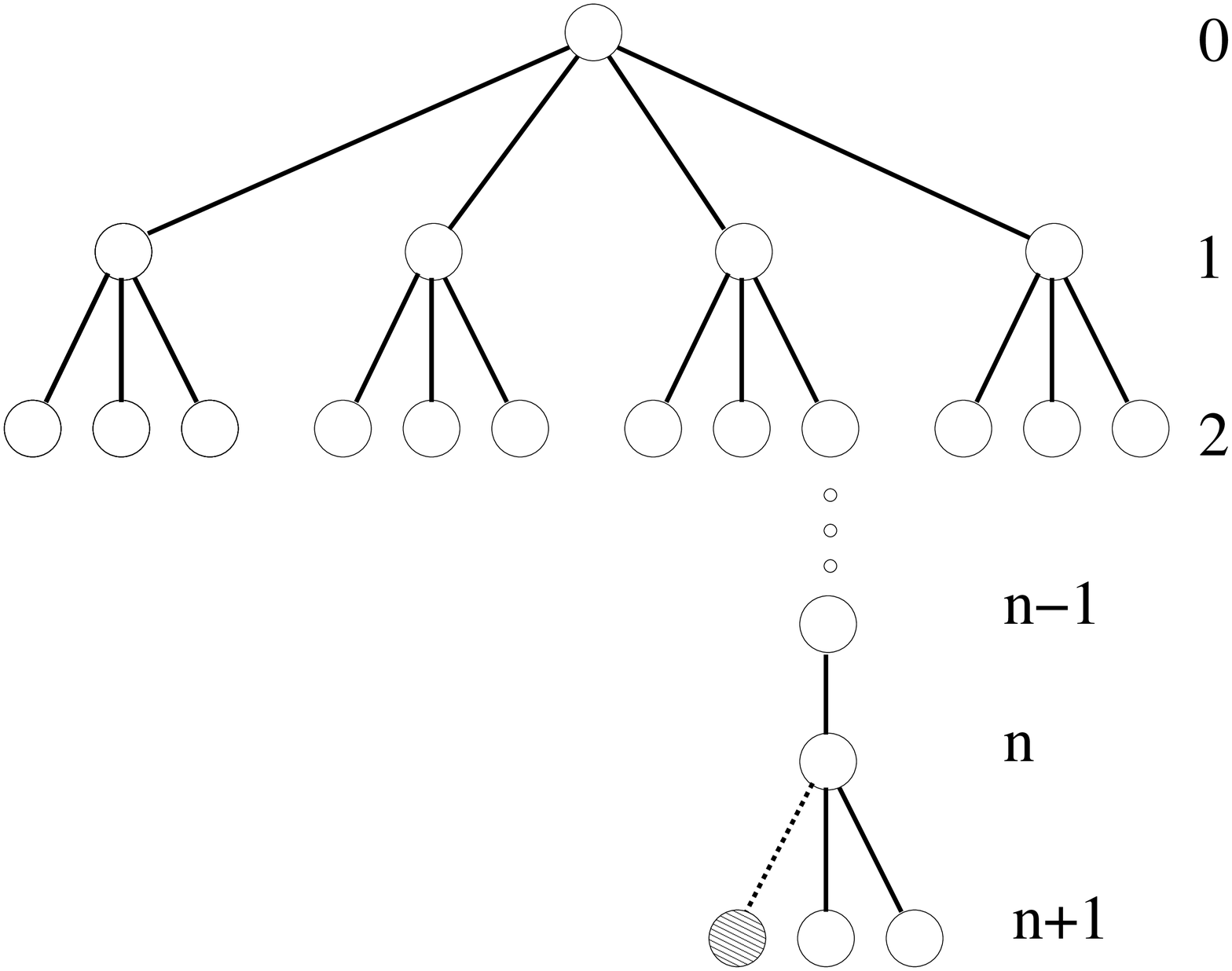} \caption{A 
Cayley tree of mixed coordination $z=3$ and $z=4$. A dark circle at 
level $n+1$ represents a $z=3$ site. Any of its four neighbors may be 
missing with probability $1/4$ but the recursion relations assume that 
the missing neighbor does not lie at level $n$. This overestimates the 
concentration of $z=4$ sites at level $n$. A fraction $c$ of $z=4$ sites 
on a random graph corresponds to a fraction $c_{eff}=4c/(3+c)$ in the 
deep interior of the Cayley tree (see text).} \label{fig4} \end{figure}

\begin{figure}[p] 
\includegraphics[width=0.5\textwidth,angle=0]{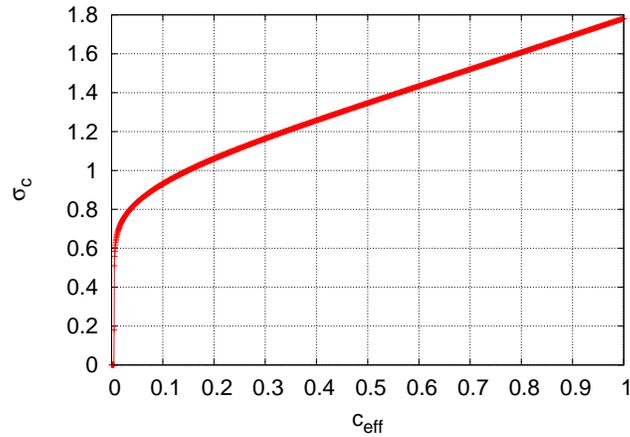} \caption{ The 
critical value $\sigma_c$ vs. $c$ as obtained from the fixed point 
equations. The concentration $c$ of $z=4$ sites on the random graph 
corresponds to $c_{eff}=4c/(3+c)$ on the $x$-axis in the above figure. 
Critical hysteresis occurs on random graphs only if $\sigma< \sigma_c$.
} \label{fig5} \end{figure}

\begin{figure}[p] 
\includegraphics[width=0.5\textwidth,angle=0]{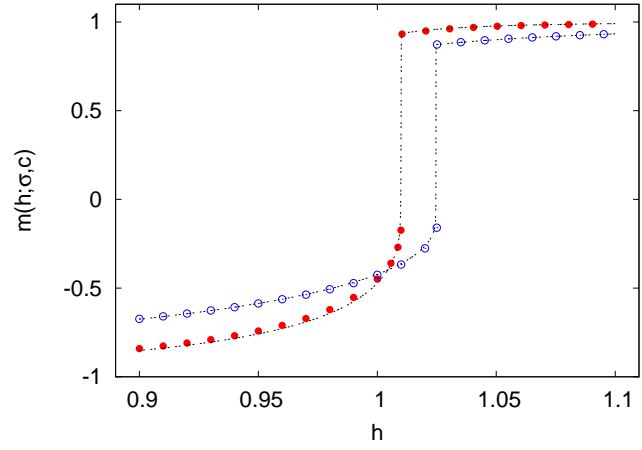} 
\caption{Magnetization $m(h;\sigma,c)$ for $c=0.1, \sigma=0.9$ 
(red/filled circles) and $c=0.9,\sigma=1.6$ (blue/open circles) near a 
discontinuity. Corresponding theoretical expressions have been 
superimposed on the simulation data for a single configuration of 
random-field for a system of size $N=10^7$.} \label{fig6} \end{figure}

\begin{figure}[p] 
\includegraphics[width=0.5\textwidth,angle=0]{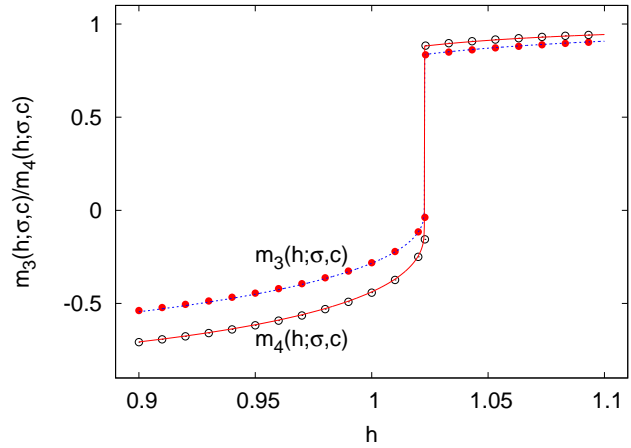} 
\caption{Magnetization $m_3(h;\sigma,c)$ and $m_4(h;\sigma,c)$ for 
$c=0.75,\sigma=1.5$ near a discontinuity. Theoretical expressions have 
been superimposed on the simulation data from a single run ($N=10^7$).} 
\label{fig7} \end{figure}

\end{document}